# Improved MDS-based Algorithm for Nodes Localization in Wireless Sensor Networks

Biljana Risteska Stojkoska [1], Vesna Kirandziska [2]

*Faculty of Computer Science and Engineering*

*University "Ss. Cyril and Methodius"*
*"Rugjer Boshkovikj" 16, 1000 Skopje, Republic of Macedonia*

[1] `biljana.stojkoska@finki.ukim.mk`
[2] `vesna.kirandziska@finki.ukim.mk`

*Abstract*— **With the recent development of technology, wireless sensor networks (WSN) are becoming an important part of many applications. Knowing the exact location of each sensor in the network is very important issue. Therefore, the localization problem is a growing field of interest. Adding GPS receivers to each sensor node is costly solution and inapplicable on nodes with limited resources. Additionally, it is not suitable for indoor environments.**

**In this paper, we propose an algorithm for nodes localization in WNS based on multidimensional scaling (MDS) technique. Our approach improves MDS by distance matrix refinement. Using extensive simulations we investigated in details our approach regarding different network topologies, various network parameters and performance issues. The results from simulations show that our improved MDS (IMDS) algorithm outperforms well known MDS-MAP algorithm [1] in terms of accuracy.**

*Keywords: wireless sensor networks, multidimensional scaling, nodes positioning*

## I. INTRODUCTION

A wireless sensor network (WSN) is a network consisting of distributed sensor devices that cooperatively monitor physical or environmental conditions at different locations. Although initially developed for military applications, today wireless sensor networks are used in many industrial and civilian application areas, including industrial process monitoring and control, machine health monitoring, environment and habitat monitoring, healthcare applications and traffic control[2][3].

After taking samples from the environment (light level, air temperature, humidity etc.) sensors process data or simply pass data through the network to a main location, known as sink node or base station. Sensors being deployed in a vast region have short range of radio communication, thus measurements cannot be sent directly to the sink node. Each sensor sends data to its closest neighbor responsible for retransmitting the packets. In term of routing, there are a lot of multi-hops protocols that offer optimal communication cost [4].

The problem of nodes localization appears in a variety of WSN applications. The information gathered from the network can often be useless if not matched with the location where it is sensed. If the sensor network is used for monitoring the temperature in a forest, nodes may be deployed from an airplane and the precise location of most sensors may be unknown. Finding the exact physical locations is a crucial issue for continual network operation and WSN management. Nodes could be equipped with a Global Positioning System (GPS), but this is a costly solution in terms of money and power consumption.

Although many different techniques have been proposed for solving sensor localization problem, it remains challenging problem in the research community [5][6]. An effective localization algorithm should use all the available information from the network to compute nodes positions. The performance of the algorithms for WSN localization depends on different network parameters, such as the network topology, the number of anchors (i.e. the anchor-to-node ratio), the radio range, the density of nodes, etc. Hence the location estimation error is going to be evaluated as a function of different parameters.

Multidimensional scaling (MDS) is a set of analytical techniques that has been used for many years in disciplines such as mathematical psychology, economics and marketing research. It is a suitable method used for reducing the dimensionality of the data, showing multidimensional data as points in two or three dimensional space [7]. This technique can also be used in WSN where only distances between nodes are known. The distance measurement between each node is used as an input data. Since MDS is a centralized technique, all measurements are collected at the sink node where further processing is done. The main advantage of using MDS is its ability to reconstruct the relative map of the network even when there are no anchor nodes (nodes with a priori known location). If given sufficient number of anchors, MDS performs very accurate position estimation enabling local map to be transformed into an absolute map.

In this paper, we investigate classical multidimensional scaling (MDS) technique for nodes localization in two dimensional WSN. We propose a variation of MDS that modifies all pairs shortest path algorithm (Dijkstra's or Floyd's) in order to decrease distance matrix error. By applying heuristic approach in distance matrix calculation we improved the accuracy compared with MDS-MAP [1]. Henceforth we will refer to our approach as Improved Multidimensional Scaling Algorithm (IMDS)

The rest of this paper is organized as follows. In the second section, the relevant work related to WSN localization techniques is discussed. The third section gives a detailed explanation of our improved MDS algorithm (IMDS). Section four gives the results provided from our simulations. Finally, we conclude this paper in section five.

## II. RELATED WORK

Many research groups have investigated different techniques for nodes localization in WSN. Most of the techniques proposed within the last years can be basically divided into two categories: range-based and range-free methods.

Range-free methods are also known as "hop-based" methods. They use hop or connectivity information for discovering nodes location [5][8]. The category of range-based methods estimates the distance between the neighboring nodes using different signal measurement techniques [9][10]. RSSI (Receive Signal Strength Indicator) is the most common technique that measures the power of the received radio signal. Other popular techniques are ToA (Time of Arrival), AoA (Angle of Arrival) and TDoA (Time Difference of Arrival). TDoA methods are very accurate, but under the Non-Line-of-Sight (NLOS) conditions their performance degenerates significantly. AoA provides more accurate result than RSSI based techniques but requires sensors equipped with additional hardware, thus appear as a more expensive solution.

Multidimensional scaling (MDS) based algorithms are ranging-based sensor localization algorithms. There are different versions of MDS for nodes localization [1][11]. The most popular is MDS-MAP, proposed by Yi Shang and Wheeler Ruml [1]. They showed that MDS-MAP outperforms other techniques, especially when applied on density networks. MDS-MAP is based on classical MDS, where the proximities of objects are treated as distances in a Euclidean space. MDS-MAP for consists of 3 steps:

1. Calculate shortest distances between every pair of nodes (using either Dijkstra's or Floyd's all pairs shortest path algorithm). This is the distance matrix that serves as input to the multidimensional scaling in step 2.

2. Apply classical multidimensional scaling to the distance matrix. The first two largest eigenvalues and eigenvectors give a relative map with relative location for each node.

3. Transform the relative map into absolute map using sufficient number of anchor nodes (at least 3).

Other approaches based on MDS exist, but they are more complex and thus more computationally dependent. Such an example is MDS-MAP(P) [12], which is a modification of the MDS-MAP based on decentralized approach. It shows better results than MDS-MAP for irregular network topologies, but requires intensive computational resources at each node. In MDS-MAP(P) each node in the network computes local map within its two-hop neighbors using MDS-MAP. Then all local maps are merged into a global map. An improvement of [12] is presented in [13], where localization is based on ordinal MDS. This variation of MDS assumes there is a linear equation which relates the shortest path distance and the Euclidean distance between each pair of nodes.

Cluster-based variations of MDS-MAP is introduced in [14][15] and investigated further in [16]. Here, the wireless network is hierarchically organized and divided into clusters. Each cluster has its own cluster head responsible for localizing its own cluster members using MDS-MAP. The positions or the nodes in each cluster are in its own coordinate system. The final step requires merging the coordinate systems of all clusters into one coordinate system. According to the simulations, this approach outperforms MDS-MAP in terms of accuracy for irregular network topology (C-shape, H-shape, S-shape, etc.) and performs smaller computational complexity compared with both MDS-MAP and MDS-MAP(P).

## III. IMPROVED MDS-BASED APPROACH FOR WSN POSITIONING

In this section we will explain in detail our improved MDS (IMDS) algorithm for nodes localization within WSN.

Classical MDS needs distance matrix of the nodes as input and produces points in two or three dimensional space (step 1 in MDS-MAP algorithm). In order to create distance matrix, the distances between each pair of nodes are needed. If there are no measured distances between some of the nodes in the network, they will be calculated using Dijkstra's shortest path algorithm. This approximation produces an error, i.e. the correct positions usually differ from the predicted ones. This error is bigger when the nodes are in multi-hop communication range.

To reduce shortest path distance estimation error, we propose an alternative approach to calculate the distance between non-neighboring nodes.

### A. Distance matrix refinement

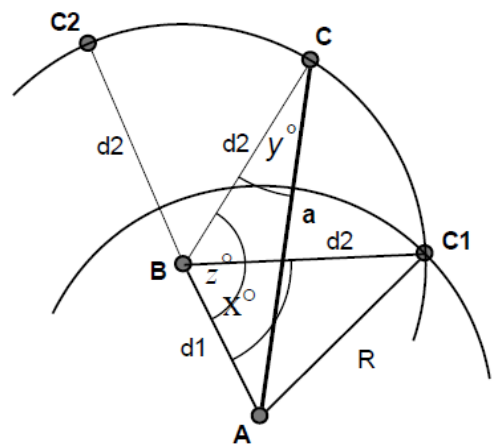

Fig. 1. Distance approximation

Consider there are three nodes in a network, (A, B and C), with known distances between nodes A and B (d1=AB), and between nodes B and C (d2=BC). Since distance matrix requires the distance between each pair of nodes in the network, the distance between nodes A and C have to be calculated.

If maximum radio range of the nodes in the network is R, than we know for sure that node C lies on the curve $C_1C_2$. If Dijkstra's algorithm is used for this purpose, it will approximate AC=AB+BC, which is the longest possible theoretical distance between nodes A and C. On the other hand, if we calculate the shortest possible theoretical distance between nodes A and C, it will corresponds to C1 (AC=R).

To minimize the possible error, we purpose heuristic solution that assumes that the node C lies exactly in the middle of the curve C1C2. Hence, the distance a=AC can be calculated using cosine formula as:

$$a^2 = d_1^2 + d_2^2 - 2 \cdot d_1 \cdot d_2 \cdot \cos(\sphericalangle ABC)$$

In order to calculate the distance *a* first we need to find the angle using cosine formula.

$$\sphericalangle ABC = \sphericalangle ABC_1 + \sphericalangle C_1BC$$

The angle $\sphericalangle ABC_1$ can be calculated again with the cosine formula:

$$\sphericalangle ABC_1 = \arccos(\frac{d_1^2 + d_2^2 - R^2}{2 \cdot d_1 \cdot d_2})$$

Since $\sphericalangle C_1BC = \sphericalangle CBC_2$,

$$\sphericalangle C_1BC = \frac{1}{2} \sphericalangle C_1BC_2$$

$$\sphericalangle C_1BC = \frac{1}{2}(\pi - \sphericalangle ABC_1)$$

$$\sphericalangle ABC = \sphericalangle ABC_1 + \frac{1}{2}(\pi - \sphericalangle ABC_1)$$

$$\sphericalangle ABC = \frac{\pi}{2} + \frac{1}{2}\sphericalangle ABC_1$$

Finally,

$$a^2 = d_1^2 + d_2^2 - 2 \cdot d_1 \cdot d_2 \cdot \cos(\sphericalangle ABC) =$$
$$= d_1^2 + d_2^2 - 2 \cdot d_1 \cdot d_2 \cdot \cos(\frac{\pi}{2} + \frac{1}{2}\sphericalangle ABC_1) =$$
$$= d_1^2 + d_2^2 + 2 \cdot d_1 \cdot d_2 \cdot \sin(\frac{1}{2}\sphericalangle ABC_1),$$

where

$$ABC_1 = \arccos(\frac{d_1^2 + d_2^2 - R^2}{2 \cdot d_1 \cdot d_2})$$

We note here that our algorithm preserves the time complexity of MDS-MAP algorithm.

## IV. SIMULATION RESULTS

### A. Network model

We assume a typical sensor network composed of hundreds of sensor nodes deployed uniformly across a monitored area. Sensors are equipped with an omni-directional antenna, hence only nodes within certain radio range R can communicate with each other. If two nodes are within each others transmission range they are called neighbors. Further, we made following assumptions:

- There is a path between every pair of nodes.
- Nodes deployed in close proximity to each other exchange messages.
- Each node uses RSSI (or any other) method for distance estimation.
- RSSI provide accurate neighboring sensor distance estimation.

We simulated and compared the results for IMDS and MDS-MAP. Our work was mainly focused on random, grid and hexagonal grid topology. Other network properties (number of anchors, average connectivity and range error) were also investigated.

We consider:
- Different network topologies:
  o random deployment (100 nodes)
  o grid deployment (100 nodes)
  o hexagonal grid deployment (100 nodes).
- Different number of anchors for absolute map construction:
  o 3, 4, 6 and 10 anchors.

In our simulation, the anchors were selected randomly.
- Different radio ranges (R) which lead to different average connectivity (average number of neighbors).
- Radio range error (from 0 to 30% of R with step 5% of R)

Thus 360 different networks were simulated (3 x 4 x 5 x 6) and each node location was discovered with both MDS-MAP and IMDS technique. The average estimation error is normalized by the radio range R:

$$Error = \frac{\sum_{i=1}^{n} \text{distance}(pos_i^{(estimated)} - pos_i^{(true)})}{(n-N) \cdot R} \cdot 100\%$$

, where n is the number of nodes in the network, N is the number of anchor nodes, $pos_i^{(estimated)}$ is the estimated location and $pos_i^{(true)}$ is the true location of the i-th node.

The connectivity parameter and the estimation error for each scenario represent average over 30 trials for both algorithms.

### B. Comparison of the MDS-MAP and IMDS

For random topologies, 100 nodes are placed randomly with a uniform distribution within a square area (10r x 10r square where r is a unit length distance). For grid and hexagonal grid topologies, 100 nodes are placed on a square grid with some placement errors, modeled as Gaussian noises. Figure 2 compares the results of MDS-MAP and IMDS for random topology with 10 anchors. Estimation errors are normalized with R, as proposed in [1][12][13]. Figure 3 and Figure 4 show the results for random topology with 10 anchors and range error of 5% of R and 10% of R respectively.

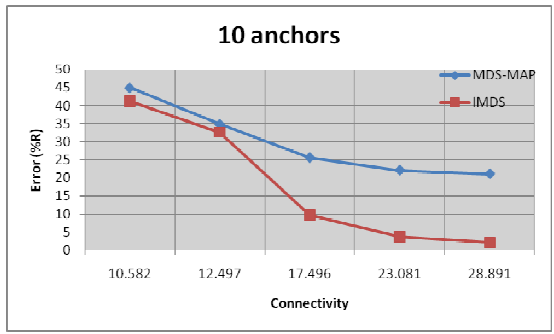

Figure 2. Comparison of the MDS-MAP and IMDS on random topology (100 nodes randomly deployed) without range error

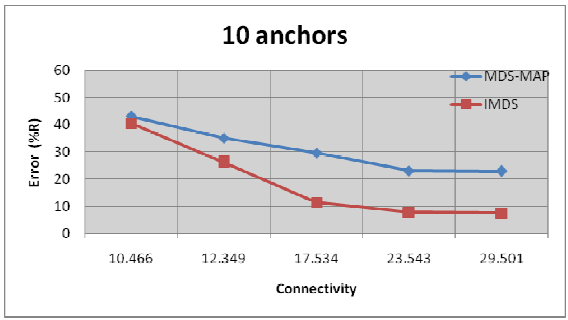

Fig. 3. Comparison of the MDS-MAP and IMDS on random topology (100 nodes randomly deployed) with range error 5% of R

As can be seen from the figures, IMDS performs smaller estimation error than MDS-MAP for both random and grid topologies.

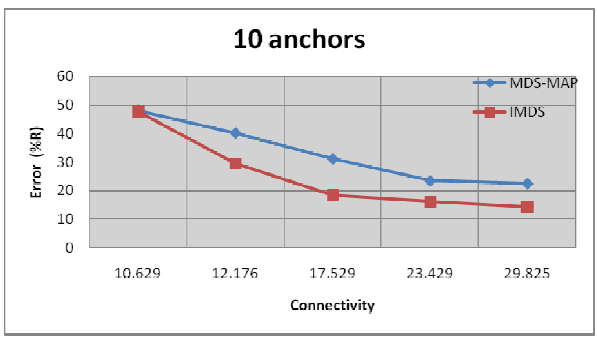

Fig. 4. Comparison of the MDS-MAP and IMDS on random topology (100 nodes randomly deployed) with range error 10% of R

Figure 5 to Figure 7 show the results for IMDS as a function of connectivity and number of anchors for random, grid and hexagonal grid topology. These networks are without range error. For grid and hexagonal grid topologies the nodes are deployed with placement error of 5%. As expected, using more anchor nodes reduces the localization error, but has no significant impact when network density is high.

Figure 8 and Figure 9 shows results of IMDS using distance measurement between neighbors with range errors from 0% to 10% of R. Having more anchors (10 vs. 6) improves performance, especially for the case with large range errors.

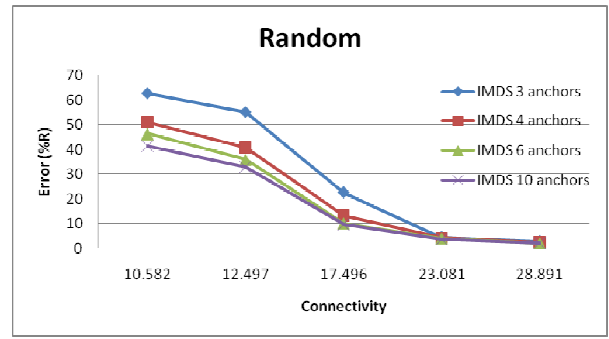

Fig. 5. The effect of number of anchors on the estimation error (100 nodes randomly deployed)

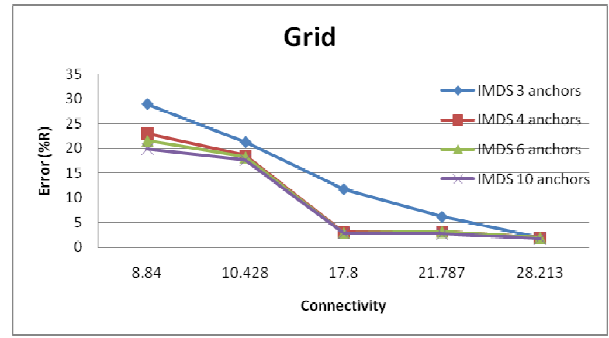

Fig. 6. The effect of number of anchors on the estimation error on grid topology (100 nodes deployed) with placement error 5%

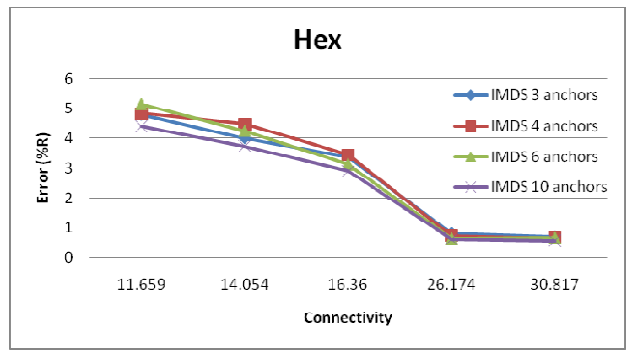

Fig. 7. The effect of number of anchors on the estimation error on hexagonal grid topology (100 nodes deployed) with placement error 5%

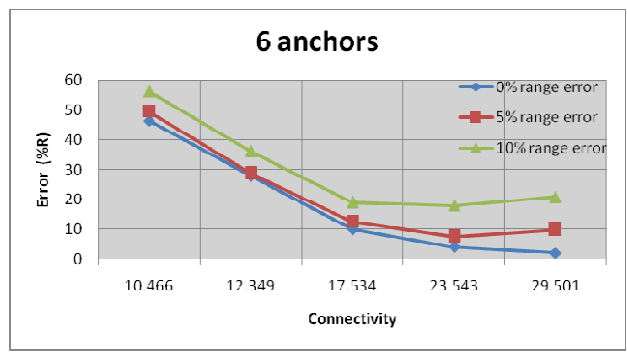

Fig. 8. The effect of radio range error on the estimation error (100 nodes randomly deployed)

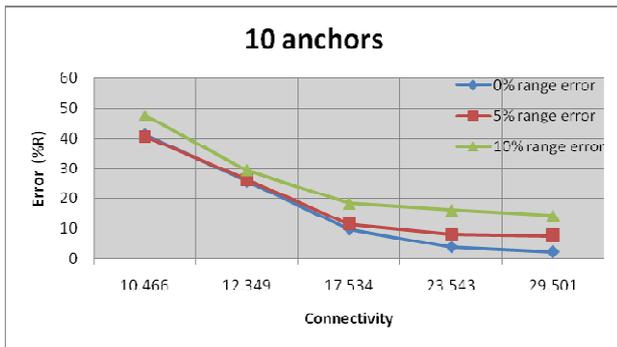

Fig. 9. The effect of radio range error on the estimation error (100 nodes randomly deployed)

## V. CONCLUSION

In this paper, we presented a new improved MDS-based algorithm (IMDS) for nodes localization in WSN. We introduce a novel technique for distance matrix refinement which should reduce the MDS error. We show that our approach outperforms well known MDS-MAP in terms of accuracy for two-dimensional networks and performs acceptable estimation error.

For future work, we intend to investigate our algorithm on irregular network topologies, considering distributed approach based on cluster formation [14]. We also plan to investigate the performance of IMDS on three-dimensional networks.